\documentclass[prl, 12pt]{revtex4}
\usepackage{bbm}
\usepackage{mathrsfs}
\usepackage{amsfonts}
\usepackage{amsmath}
\usepackage{times}
\usepackage{graphicx}
\usepackage{subfigure}
\usepackage{bm}
\begin{document}
\title{Quantum Geometric Tensor (Fubini-Study Metric) in Simple Quantum System: A pedagogical Introduction}
\author{Ran Cheng}
\affiliation{Department of Physics, University of Texas, Austin, TX 78712, USA}
\email{rancheng@physics.utexas.edu}
\date{\today}

\begin{abstract}
Geometric Quantum Mechanics is a novel and prospecting approach motivated by the belief that our world is ultimately geometrical. At the heart of that is a quantity called Quantum Geometric Tensor (or Fubini-Study metric), which is a complex tensor with the real part serving as the Riemannian metric that measures the `quantum distance', and the imaginary part being the Berry curvature. Following a physical introduction of the basic formalism, we illustrate its physical significance in both the adiabatic and non-adiabatic systems.
\end{abstract}
\maketitle

\section{Introduction}

The most intriguing feature of modern physics is the introduction of geometrical concepts describing fundamental principles of nature~\cite{ref:KHuang}. One one hand, the gravity emerges as the local space-time symmetry, where comparison between nearby local frames naturally gives rise to the concept of Christoffel connection; On the other hand, electroweak and strong interactions are unified by Yang-Mills theory, which identifies the gauge interactions as local symmetries of internal degrees of freedom. Similarly, comparison between nearby frames of the internal spaces (e.g., for SU(2), the three isospin axis) introduces the gauge connection. In electromagnetic theory, this reduces to the Weyl's principle where the gauge connection is the usual four-potential $\mathcal{A}_\mu$. The nature resumes all the observed interactions by simply obeying space-time and gauge symmetries.

In early 1980's, people discovered that gauge fields not only appear in fundamental forces between elementary particles, they also emerge in simple quantum systems under certain constrains~\cite{ref:BerryPhase, ref:WZ, ref:AAPhase, ref:NonAbelianBO}. For example, when a spin-$1/2$ electron adiabatically follows a smoothly varying magnetization texture, an effective gauge field now known as Berry curvature affects the motion of the electron. Following the same line, people further found that this interesting gauge structure is just the Holonomy effect on the phase bundle of the wave function~\cite{ref:AAGeomQM}. Specifically, when we identify states differ only by a local phase factor (since physical observable is blind to the phase), the Hilbert space $\mathcal{H}$ reduces to the Projected Hilbert space $\mathcal{PH}$ and quantum states become `Rays'. In this particular space, people were able to construct a geometric reformulation of the usual Schrodinger quantum mechanics~\cite{ref:AAGeomQM, ref:Tze, ref:notes}, where covariant derivative is enabled by the emergent gauge potential. Studies along this path is often named geometric quantum mechanics~\cite{ref:RevGeoQM}.

What is more significant in geometric quantum mechanics is the emergent metric structure in addition to the gauge structure mentioned above. Historically, the discovery of this metric structure precedes the intensive study on emergent gauge fields in an attempt to define `quantum distance' (or interval) between different states~\cite{ref:QGT}. A remarkable feature of the quantum distance comes from the fact that quantum states are denoted by complex functions, which renders the metric a complex tensor known as the Quantum Geometric Tensor (QGT). Thanks to the Hermitean property of the inner product in quantum mechanics, the real and imaginary parts of this complex tensor play quite separate roles. The former is symmetric and serves as the Riemannian tensor fulfilling the function of measuring quantum distance, while the latter is antisymmetric and is identified with the emergent gauge field in projected Hilbert space. There is a simple way to see this in advance: when taking the inner product between two quantum states we need two information, the overlap and the relative phase. The symmetric part of QGT measures the former, while the antisymmetric part gives the flux density of the latter.

If we represent a state $|\psi\rangle=\sum_k Z_k |e_k\rangle$ by the complex vector $\mathbb{Z}:=[Z_1, Z_2, \cdots, Z_n]$ where $|e_k\rangle$ is a set of orthonormal basis, the QGT is nothing but the Fubini-Study metric (FSM) on the $CP^n$ manifold~\cite{ref:EvolveFS}. To avoid confusion in terminology, we would prefer QGT in the following discussions. The QGT has enjoying renewed interests in the study of quantum statistical mechanics~\cite{ref:StatDist}, quantum transports in solids~\cite{ref:Vanderbilt}, quantum phase transitions~\cite{ref:Zanardi}, topological insulators~\cite{ref:MatsuuraRyu}, \emph{etc.}. In the following sections, we are going to give a more physical introduction to the basic formalism of QGT and its implications in both adiabatic and non-adiabatic systems.

\section{Formalism - A Physicist's way}
Let us consider a family of parameter-dependent Hamiltonian~\cite{ref:QGT, ref:MatsuuraRyu, ref:YQMa, ref:RanNiu} $\{H(\lambda)\}$ requiring a smooth dependence on a set of parameters $\lambda=(\lambda_1,\lambda_2,...)\in \mathcal{M}$, which consists of the base manifold of the quantum system. The Hamiltonian acts on the parameterized Hilbert space $\mathcal{H}(\lambda)$, the eigen-energies and eigen-states are denoted by $E_n(\lambda)$ and $|\phi_n(\lambda)\rangle$ respectively. The system state $|\psi(\lambda)\rangle$ is a linear combination of $|\phi_n(\lambda)\rangle$ at each point in $\mathcal{M}$.

Upon infinitesimal variation of the parameter $\mathrm{d}\lambda$, we define the quantum distance:
\begin{align}
\mathrm{d}s^2=||\psi(\lambda+\mathrm{d}\lambda)-\psi(\lambda)||^2=\langle\delta\psi|\delta\psi\rangle&=\langle\partial_\mu\psi|\partial_\nu\psi\rangle\mathrm{d}\lambda^\mu\mathrm{d}\lambda^\nu \notag\\ &=(\gamma_{\mu\nu}+i\sigma_{\mu\nu})\mathrm{d}\lambda^\mu\mathrm{d}\lambda^\nu
\end{align}
where in the last line we have decomposed the complex tensor $\langle\partial_\mu\psi|\partial_\nu\psi\rangle$ by its real and imaginary parts. Since the inner product of any two states is Hermitean, we know that $\gamma_{\mu\nu}+i\sigma_{\mu\nu}=\gamma_{\nu\mu}-i\sigma_{\nu\mu}$, which indicates the symmetric properties of the two tensors:
\begin{align}
\gamma_{\mu\nu}&=\gamma_{\nu\mu} \notag\\
\sigma_{\mu\nu}&=-\sigma_{\nu\mu}
\end{align}
so that $\sigma_{\mu\nu}\mathrm{d}\lambda^\mu\mathrm{d}\lambda^\nu$ vanishes due to the antisymmetry of $\sigma_{\mu\nu}$ and symmetry of $\mathrm{d}\lambda^\mu\mathrm{d}\lambda^\nu$, thus the quantum distance reduces to $\mathrm{d}s^2=\langle\delta\psi|\delta\psi\rangle=\gamma_{\mu\nu}\mathrm{d}\lambda^\mu\mathrm{d}\lambda^\nu$.

However, a careful look reminds us that $\gamma_{\mu\nu}$ thus defined is NOT gauge invariant which disqualifies this tensor as the appropriate metric on measuring the quantum distance. Specifically, when we take $|\psi'(\lambda)\rangle=\exp^{i\alpha(\lambda)}|\psi(\lambda)\rangle$ and define $\langle\partial_\mu\psi'|\partial_\nu\psi'\rangle=\gamma'_{\mu\nu}+i\sigma'_{\mu\nu}$, a simple calculation shows that:
\begin{align}
\gamma'_{\mu\nu}&=\gamma_{\mu\nu}-\beta_\mu\partial_\nu\alpha-\beta_\nu\partial_\mu\alpha+\partial_\mu\alpha\partial_\nu\alpha \notag\\
\sigma'_{\mu\nu}&=\sigma_{\mu\nu}
\end{align}
where $\beta_\mu(\lambda)=i\langle\psi(\lambda)|\partial_\mu\psi(\lambda)\rangle$ is the Berry connection~\cite{ref:BerryPhase}, which is purely real due to the normalization $\langle\psi(\lambda)|\psi(\lambda)\rangle=1$. It is obvious that upon the above gauge transformation, the Berry connection changes as $\beta'_\mu=\beta_\mu+\partial_\mu\alpha$. Thus a rescue is to redefine the gauge invariant metric:
\begin{align}
g_{\mu\nu}(\lambda):=\gamma_{\mu\nu}(\lambda)-\beta_\mu(\lambda)\beta_\nu(\lambda)
\end{align}
where changes from the second term on the right hand side cancels the changes from the first part under a gauge transformation, so that $g'_{\mu\nu}(\lambda)=g_{\mu\nu}(\lambda)$. We can understand this relation through a physicist's way, $\gamma_{\mu\nu}$ measures the distance of `bare states' in Hilbert space $\mathcal{H}$, while $g_{\mu\nu}$ measures the distance of `Rays' in Projected Hilbert space $\mathcal{PH}=\mathcal{H}/U(1)$. But physical observable relate to Hermitean operators acting on `Rays', not the `bare states', required by the principle of gauge invariance. Therefore, it is safe to discard $\gamma_{\mu\nu}$ and focus on the gauge invariant metric $g_{\mu\nu}$ in $\mathcal{PH}$ in the following discussions. For mathematical unambiguity and simplicity, we further define the `Quantum Geometric Tensor' (QGT), which is the Fubini-Study metric on quantum Rays, as:
\begin{align}
Q_{\mu\nu}(\lambda):=\langle\partial_\mu\psi(\lambda)|\partial_\nu\psi(\lambda)\rangle-\langle\partial_\mu\psi(\lambda)|\psi(\lambda)\rangle\langle\psi(\lambda)|\partial_\nu\psi(\lambda)\rangle
\end{align}
thus quantities previously defined are related to QGT by,
\begin{align}
g_{\mu\nu}=\mathrm{Re}\ Q_{\mu\nu}; \qquad  \sigma_{\mu\nu}=\mathrm{Im}\ Q_{\mu\nu}
\end{align}

To see more explicitly that the gauge invariant tensor $g_{\mu\nu}$ indeed plays the role of a metric, we now turn to a different approach which comes from perturbation theory. Take the inner product of the state $|\psi(\lambda)\rangle$ with $|\psi(\lambda+\mathrm{d}\lambda)\rangle$ up to second order in $\mathrm{d}\lambda$,
\begin{align}
\langle\psi(\lambda)|\psi(\lambda+\mathrm{d}\lambda)\rangle=1+i\beta_\mu(\lambda)\mathrm{d}\lambda^\mu+\frac12\langle\psi(\lambda)|\partial_\mu\partial_\nu\psi(\lambda)\rangle\mathrm{d}\lambda^\mu\mathrm{d}\lambda^\nu \label{eq:innerproduct}
\end{align}
since $\langle\psi|\partial_\mu\psi\rangle\in \mathrm{Im}$ (pure imaginary), we know that $\langle\partial_\mu\psi|\partial_\nu\psi\rangle+\langle\psi|\partial_\mu\partial_\nu\psi\rangle\in \mathrm{Im}$ so that $\mathrm{Re}\langle\psi|\partial_\mu\partial_\nu\psi\rangle=-\mathrm{Re}\langle\partial_\mu\psi|\partial_\nu\psi\rangle=-\gamma_{\mu\nu}$, thus we obtain from Eq.~\eqref{eq:innerproduct} the gauge invariant result:
\begin{align}
|\langle\psi(\lambda)|\psi(\lambda+\mathrm{d}\lambda)\rangle|&=1-\frac12(\gamma_{\mu\nu}(\lambda)-\beta_\mu(\lambda)\beta_\nu(\lambda))\mathrm{d}\lambda^\mu\mathrm{d}\lambda^\nu\notag\\
&=1-\frac12 g_{\mu\nu}(\lambda)\mathrm{d}\lambda^\mu\mathrm{d}\lambda^\nu \label{eq:distance}
\end{align}
for two quantum states labeled by $\lambda_I$ and $\lambda_F$, the quantum distance between them is therefore expressed as the integration over the metric:
\begin{align}
|\langle\psi(\lambda_F)|\psi(\lambda_I)\rangle|=1-\frac12\int_{\lambda_I}^{\lambda_F} g_{\mu\nu}(\lambda)\mathrm{d}\lambda^\mu\mathrm{d}\lambda^\nu \label{eq:finitedist}
\end{align}
the last term in this equation is the length of the geodesic curve marked by the metric $g_{\mu\nu}$ and we call it `geodesic quantum distance'. It is worthy of noticing that the inner product of any two states should within the range of $[0,1]$, by which we regard the QGT as a metric measuring the geodesic distance of points lying on the Bloch sphere. Specifically, if we define $|\langle\psi|\chi\rangle|=\cos^2\frac{\theta}2$, then $\mathrm{d}\theta=2\mathrm{d}s=2\sqrt{|g_{\mu\nu}\mathrm{d}\lambda^\mu\mathrm{d}\lambda^\nu|}$.

\section{Case One -- Adiabatic System}

If a quantum system is confined on a single energy level, where transitions to other levels are negligible due to large energy gaps separating this particular level, the QGT can be defined with the instantaneous eigenstate of that level. It losses no generality to consider the ground state as an example where the energy is denoted by $E_0$, and the corresponding eigenstate is labeled by $|\phi_0(\lambda)\rangle$. We assume a sufficiently large energy gap `protecting' the ground state thus transitions to any excited states are ignored. In condensed matter physics, people often interpret such a situation as the absence of Goldstone modes and identify the elementary excitations as `massive'. This is a very common case in the study of Fractional Quantum Hall liquid, High Tc superconductivity, and the recent progress on quantum phase transitions~\cite{ref:Zanardi}. However, it is still far from clear whether there exists an intrinsic relationship between the real and imaginary parts of the QGT near the critical point of the transition~\cite{ref:RanNiu}.

The instantaneous ground eigenstate of the system is defined as $\hat{H}(\lambda)|\phi_0(\lambda)\rangle=E_0(\lambda)|\phi_0(\lambda)\rangle$, the adiabaticity guarantees the restriction of the system to the subspace $\mathcal{H}_{E_0}(\lambda)$ of the Hilbert space. First we assume the ground state is non-degenerate, take the partial derivative $\partial_\mu$ on both sides of the above relation and consider the orthonormal condition $\langle\phi_n(\lambda)|\phi_0(\lambda)\rangle=\delta_{n0}$, we arrive at the Feynman-Hellman equations:
\begin{align}
&\langle\phi_n|\phi_0\rangle=\frac{\langle\phi_n|\partial_\mu H|\phi_0\rangle}{E_0-E_n}\quad \mbox{if}\ n\neq 0 \notag\\
&\langle\phi_0|\partial_\mu H|\phi_0\rangle=\partial_\mu E_0 \label{eq:FeynmanHelleman}
\end{align}
Then the QGT can be defined on the ground state as:
\begin{align}
Q_{\mu\nu}&=\langle\partial_\mu\phi_0|(\mathbf{1}-|\phi_0\rangle\langle\phi_0|)|\partial_\nu\phi_0\rangle \notag\\
&=\sum_{n\neq0}\langle\partial_\mu\phi_0|\phi_n\rangle\langle\phi_n|\partial_\nu\phi_0\rangle \notag\\
&=\sum_{n\neq0}\frac{\langle\phi_0|\partial_\mu H|\phi_n\rangle\langle\phi_n|\partial_\nu H|\phi_0\rangle}{(E_0-E_n)^2} \label{eq:sumform}
\end{align}
its real part $g_{\mu\nu}=\mathrm{Re}Q_{\mu\nu}$ is the Riemann metric introduced in the former section which relates directly to the `Fidelity Susceptibility' in the study of quantum phase transition~\cite{ref:Zanardi}, and its imaginary part $\sigma_{\mu\nu}=\mathrm{Im}\ Q_{\mu\nu}$ only differs from the Berry curvature~\cite{ref:BerryPhase, ref:WZ} by a factor of $-2$ because:
\begin{align}
F_{\mu\nu}=\partial_{[\mu,}\beta_{\nu]}=i\langle\partial_{[\mu}\phi_0|\partial_{\nu]}\phi_0\rangle=i(Q_{\mu\nu}-Q_{\nu\mu})=-2\mathrm{Im}Q_{\mu\nu}=-2\sigma_{\mu\nu}
\end{align}
Therefore, we finally arrive at the relation:
\begin{align}
Q_{\mu\nu}=g_{\mu\nu}-\frac i2F_{\mu\nu} \label{eq:corerelation}
\end{align}
It seems that Eq.~\eqref{eq:sumform} provides no better way to calculate the QGT than the simple algorithm of taking partial derivatives on the eigenstates $|\phi_n(\lambda)\rangle$. However, in a real quantum system, the Hamiltonian is usually too complicated to be solved analytically and people usually have to resort to numerical solution of the eigenstates. But the phase relations between two neighboring sets of solutions $|\phi_n(\lambda_i)\rangle$ and $|\phi_n(\lambda_{i+1})\rangle$ are completely random in computer program, thus it losses sense to take partial derivative $|\partial_\mu\phi_n(\lambda)\rangle$ as it will never give a definite value. Eq.~\eqref{eq:sumform} removes this phase ambiguity by transforming the partial derivative of the eigenstates to that of the Hamiltonian, which is always well defined. Moreover, the form is explicitly gauge invariant and the term $(E_0-E_n)^2$ in the denominator implies the singular behavior of the QGT near degenerate points.

If the ground state is degenerate $\hat{H}(\lambda)|\phi_{0i}(\lambda)\rangle=E_0(\lambda)|\phi_{0i}(\lambda)\rangle$ where $i$ labels the second quantum number, the QGT becomes a matrix with non-Abelian features~\cite{ref:NonAbelianBO, ref:YQMa, ref:RanNiu}. By generalizing the Feynman-Hellman relations Eq.~\eqref{eq:FeynmanHelleman} to degenerate ground states, we obtain the modified expression:
\begin{align}
[Q_{\mu\nu}]_{ij}=\sum_{n\neq0, k(n)}\frac{\langle\phi_{0i}|\partial_\mu H|\phi_{nk}\rangle\langle\phi_{nk}|\partial_\nu H|\phi_{0j}\rangle}{(E_0-E_n)^2}
\end{align}
where $k(n)$ labels the possible degeneracy of the $n$-th level. But we won't go into the physical applications of the non-Abelian QGT here, ambitious readers are highly recommended to read Ref.~\cite{ref:YQMa, ref:RanNiu}.

A typical example of the adiabatic case is the spin-$\frac12$ particle subject to a magnetic field with constant amplitude and slowly time-varying orientation. When the frequency of the variation of the magnetic field is much smaller than the zeeman energy of the spin, we may assume the adiabatic condition that the spin always follows the instantaneous direction of the magnetic field. The Hamiltonian reads $H=\mu\vec{\sigma}\cdot\vec{B}$ where $\mu$ is the Gyromagnetic ratio constant, it is straightforward to solve the two eigenstates:
\begin{align}
|+\rangle=\left(e^{-i\phi/2}\cos\frac\theta2,\ e^{i\phi/2}\sin\frac\theta2\right)^T,\qquad |-\rangle=\left(-e^{-i\phi/2}\sin\frac\theta2,\ e^{i\phi/2}\cos\frac\theta2\right)^T
\end{align}
where $\theta$ and $\phi$ are spherical angles specifying the direction of magnetic field. Using Eq.~\eqref{eq:sumform}, we obtain the Riemannian metric and the Berry curvature ($\hbar=1$ for simplicity):
\begin{align}
g_{\mu\nu}=\left(
               \begin{array}{cc}
               1 & 0 \\
               0 & \sin^2\theta
               \end{array}
           \right),
\qquad
F_{\mu\nu}=\frac12\left(
                      \begin{array}{cc}
                      0 & -\sin\theta \\
                      \sin\theta & 0
                      \end{array}
\right)
\end{align}
where $\mu$ and $\nu$ run between $\theta$ and $\phi$. The Riemannian tensor is just the metric on $S^2$, namely, the Bloch sphere of the spin wave function; the Berry curvature is a symplectic form on $S^2$ and if we transform it to the usual magnetic field $\mathbb{B}^r=\frac12\frac{\epsilon^{r\mu\nu}}{r^2\sin\theta}F_{\mu\nu}=\frac1{2r^2}$, it turns out to be the magnetic field originating from a monopole located at the origin with magnetic charge $1/2$.

\section{Case Two -- non-Adiabatic System}

Now, we release the adiabatic condition and turn to the most general case of an arbitrary quantum evolution governed by an arbitrary Hamiltonian $H(t)$, it may be non-linear, non-periodic, no parametrical dependence on $\lambda$, and may in general be time-dependent.

This time, we are going to register the evolution of the state by time $t$ rather than the parameter $\lambda$. Expand $|\psi(t+\mathrm{d}t)\rangle$ to second order in $\mathrm{d}t$,
\begin{align}
|\psi(t+\mathrm{d}t)\rangle=|\psi(t)\rangle+\frac{\mathrm{d}}{\mathrm{d}t}|\psi(t)\rangle\mathrm{d}t+\frac12\frac{\mathrm{d^2}}{\mathrm{d}t^2}|\psi(t)\rangle\mathrm{d}t^2+\cdots
\end{align}
meanwhile, by full Schrodinger equation we also know that,
\begin{align}
\frac{\mathrm{d}}{\mathrm{d}t}|\psi(t)\rangle=-\frac i\hbar H(t)|\psi(t)\rangle,\qquad \frac{\mathrm{d^2}}{\mathrm{d}t^2}|\psi(t)\rangle=-\frac i\hbar \frac{\mathrm{d}H(t)}{\mathrm{d}t}|\psi(t)\rangle-\frac1{\hbar^2}H(t)^2|\psi(t)\rangle
\end{align}
Before checking the inner product of neighboring states, we should note an important quantity -- the energy uncertainty (or energy fluctuation) defined as $(\Delta E)^2=\langle\psi|H^2|\psi\rangle-\langle\psi|H|\psi\rangle^2$. If we go back to the adiabatic case, this quantity would be zero where the system has a determined energy; but for a general process, it is non-trivial and may also be a function of time $\Delta E=\Delta E(t)$. In view of all the above relations, some manipulations lead us to the following equation:
\begin{align}
|\langle\psi(t)|\psi(t+\mathrm{d}t)\rangle|=1-\frac12\frac{(\Delta E)^2}{\hbar^2}\mathrm{d}t^2 +\mathcal{O}(\mathrm{d}t^4) \label{eq:timesecondorder}
\end{align}
Remember we have defined the quantum distance between two arbitrary states by an `angle' $\theta$ on the Bloch sphere as $|\langle\psi|\chi\rangle|=\cos^2\frac{\theta}2$, then we immediately obtain from Eq.~\eqref{eq:timesecondorder} an interesting relation:
\begin{align}
\frac{\mathrm{d}\theta}{\mathrm{d}t}=\frac{2|\Delta E|}\hbar \qquad \mbox{or} \qquad \theta=2\int\frac{|\Delta E|}\hbar \mathrm{d}t\ \in\ [0,\pi] \label{eq:AAFormular}
\end{align}
The term $\frac{\mathrm{d}\theta}{\mathrm{d}t}$ has an obvious physical interpretation --  the `quantum velocity', i.e., the evolution rate of a quantum state. Eq.~\eqref{eq:AAFormular} relates the quantum velocity to the energy uncertainty of that system: the larger the fluctuation in energy, the faster the quantum evolution. \emph{It is the energy fluctuation that drives the quantum evolution of a system.} This is a crucial discovery in fundamental quantum theory two decades ago now known as the famous `Anandan-Aharonov theorem'~\cite{ref:AAGeomQM}.

What is the relationship between the Anandan-Aharonov theorem and the QGT? Let us retrieve the parameter $\lambda$ but this time it is not necessarily an adiabatic parameter. We have seen the geodesic quantum distance $\mathrm{d}\theta=2\sqrt{|g_{\mu\nu}\mathrm{d}\lambda^\mu\mathrm{d}\lambda^\nu|}$ in the previous sections, taking into account Eq.~\eqref{eq:AAFormular} we would obtain,
\begin{align}
|\Delta E|=\hbar\sqrt{|g_{\mu\nu}\dot{\lambda}^\mu\dot{\lambda}^\nu|} \qquad \mbox{(\emph{dot denotes time derivative})}
\end{align}
which gives us another way to justify how well the adiabaticity is hold: The slower the parameter varies with time, the smaller the energy uncertainty, i.e., the system is maintained in a single energy level.

Strictly speaking, when a system is far from the adiabatic region, the QGT defined on the overall state $|\psi\rangle$ is totally different from that defined on a particular eigenstate. Among the physics community, people prefer to call the former the Fubini-Study metric and the latter simply Quantum Geometric Tensor. Of course, this is just a matter of terminology, and the two tend to be equivalent at the adiabatic limit. We will not attempt to distinguish them in this literature. A specific illustration of this subtle difference is provided in Ref.~\cite{ref:RanNiu2} on a similar spin-$1/2$ model discussed above but the variation of the magnetic field there is non-adiabatic. The QGT there is the same as the adiabatic case while the FSM , though has the same form, depends on a different set of spherical angles.

\section{Summary}

In this article, we introduced the concept of Quantum Geometric Tensor, which is the Fubini-Study metric furnishing the phase bundle of a quantum system. It gives rise to the geodesic quantum distance defined on the Bloch sphere of a quantum state measured by the Riemannian metric, which composes the real part of this tensor. In a general quantum evolution, the metric form also determines the rate of change of the system thus defining the quantum velocity. Meanwhile, the imaginary part of this tensor which is antisymmetric, plays the role of the Berry curvature, the integral of which gives the gauge invariant geometric phase of the wave function.\\

\emph{Acknowledgements} - The author is grateful for Prof. Qian Niu, X. Li, and Y.-Z. You for helpful discussions. Special thank is also given to the Seminar of Quantum Field Theory held by Tsinghua University, which invited me to give a talk on this topic on summer 2010.


\begin{thebibliography}{}
\bibitem{ref:notes} For relativisitc quantum mechanic, the geometrical reformulation is still absent today. This is partly due to the replacement of particles by field quanta, but quantum field theory already plays the role of a geometrical theory.
\end{thebibliography}
\end{document}